\documentclass[epj]{svjour}

\usepackage{graphicx}
\usepackage{fancyhdr}

\def\mytitle{My title} 
\def\myauthors{My name}  
\def\mytype{My type of session}
\def\mysession{My session}

\def\mytitle{Rare B decays at LHCb} 
\def\myauthors{Michela Lenzi}    
\def\mytype{Contributed Talk}    
\def\mysession{Flavor Physics}

\begin{document}
\title{Rare B decays at LHCb}
\author{Michela Lenzi\inst{1} 
\thanks{\emph{Email:} Michela.Lenzi@fi.infn.it}%
\\On behalf of LHCb collaboration
}                     
%
%
\institute{INFN sezione di Firenze, 
via G. Sansone 1, 50019 Sesto Fiorentino, Firenze, Italy
}
%

\date{}
\abstract{
%
Rare loop-induced decays are sensitive to New Physics in many Standard Model extensions. In this 
paper we discuss the potential of the LHCb experiment to 
very rare $\mathrm{B_s} \rightarrow \mu^+ \mu^-$ decays, radiative penguin   
$\mathrm{b \rightarrow s}\gamma$ decays and electroweak penguin $\mathrm{b \rightarrow s\ell\ell}$ decays.
The experimental strategies and the expected sensitivities are presented.
} 
\maketitle
\section{Introduction}
\label{intro}

Although the Standard Model (SM) is successful
in explaining almost all experimental results
of elementary particle physics, it is possible
that physics beyond the SM exists just above the
presently reachable energy scale.
New Physics is expected to be accessible from rare 
decays where standard model contributions are suppressed enough to allow 
potential small effects to emerge.
In this paper we
will focus on the flavor-changing neutral currents
(FCNCs) processes, 
which, in the Standard Model, are forbidden at the tree level and can
proceed only via loop diagrams. 
If additional box and/or penguin diagrams
with non-SM particles contribute to these
processes, the complex couplings of new particles may result in an
enhancement of decay rates 
or in the appearance of non-trivial CP-violating phases.

The LHC will be a copious source of B mesons,
with a total $\mathrm{b\bar{b}}$ cross-section of $\sim$500 $\mu b$.
LHCb is a forward spectrometer 
for b physics. Its main features are a precise
vertex detector, two RICH detectors and a
versatile trigger with a 2 kHz output rate dominated
by $\mathrm{b\bar{b}}$ events. LHCb will operate
at a luminosity of $\mathcal{L} = 2 \times 10^{32} cm^2s^{-1}$,
corresponding to 2 $fb^{-1}$ per year.

The reconstruction of rare b decays at LHC is
a challenge due to the small rates and large backgrounds
from various sources. The most critical
is the combinatorial background from pp$\rightarrow$bbX
events, containing secondary vertices and characterized
by high charged and neutral multiplicities.
The studies reported in this paper have been performed using 
fully simulated events.

\section{Very rare decay $\mathrm{B^0_s} \rightarrow \mu\mu $}
\label{sec:1} 

Given its simple experimental signature and the 
clear theoretical picture for its prediction,
the measurement of the BR of the very rare decay
$\mathrm{B^0_s} \rightarrow \mu^+\mu^-$ 
is an excellent probe of New Physics effects.
This FCNC process is also helicity suppressed, 
resulting in a prediction for the SM branching ratio of 
$(3.55 \pm 0.33) \times 10^{-9}$. 
This branching ratio is known to increase as the sixth power of the
ratio of the Higgs vacuum in MSSM
expectation value, tan$\beta$ \cite{tanb}.
Any improvement on the limit on this BR is therefore particularly 
important to probe large-tan$\beta$ models.
The anomalus muon magnetic moment measured at BNL disagrees with the SM
expectation by 
2.7$\sigma$ 
In the context of CMSSM 
at large tan$\beta$ ($\sim$ 50), this indicates that the gaugino 
mass is in the range 400-650 GeV/$c^2$, that corresponds to a
$\mathrm{B^0_s} \rightarrow \mu^+\mu^-$
 branching ratio of $10^{-7}-10^{-9}$. 
The present limit on the BR provided by Tevatron is $< 7.5 \times
10^{-8}$ at 90\% CL, that is expected 
to improve up to $<2.0 \times 10^{-8}$ by the end of the Tevatron
run. 
This is still about 6 times higher than the SM expected value.
Given the extremely low branching ratio of
the signal, a detailed understanding of the
background is crucial in this analysis. 
Several sources of background were considered:
combinatorial background (two
real muons that combine to form a
signal candidate); misidentified hadrons and  
exclusive decays with very small branching
ratios, that could simulate the signal.

A good invariant mass resolution is crucial to
reduce the combinatorial
background, but also to reduce the contamination of
misidentified two-body decays. Good mass
resolution also allows a clear separation
between  $\mathrm{B_d}$ and $\mathrm{B_s}$ decays. 
\begin{figure*}[htb]
\centering
\includegraphics[width=0.45\textwidth,height=7.5cm,angle=0]{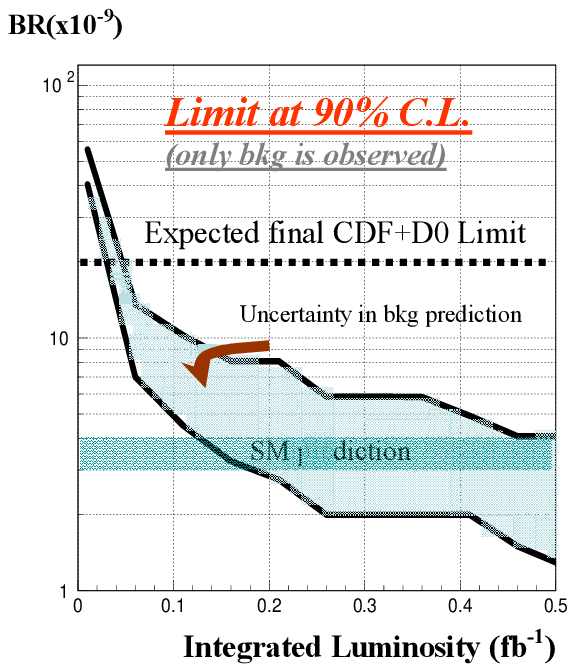}
\includegraphics[width=0.45\textwidth,angle=0]{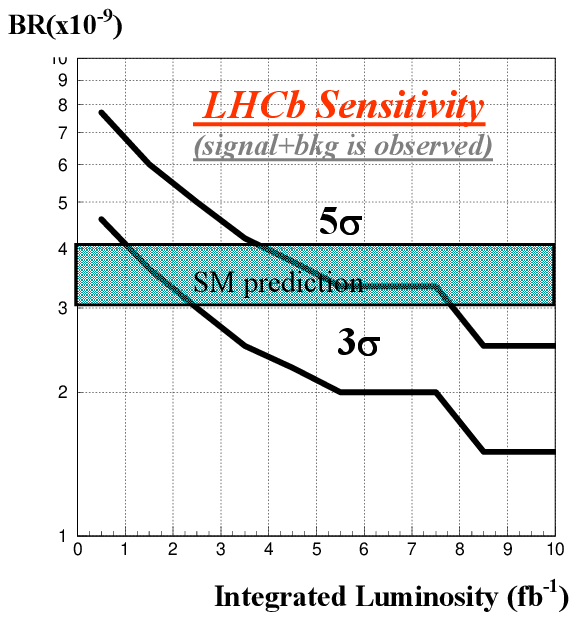}
\caption{Left: BR($\times 10^{-9}$) exclusion at 90\% confidence level
 as a function of the integrated luminosity (fb$^{-1}$), under the
 assumption that only background is present.
Right: BR($\times 10^{-9}$) observation ($3\sigma$) or discovery ($5\sigma$)
as a function of the integrated luminosity (fb$^{-1}$).
}
\label{confidencelevel}       
\label{sensitivity}       
\end{figure*}
Good muon identification and good 
vertex resolution are also critical.
The LHCb two-track vertex resolution is
$\sim 110 \mu$m in the $\phi$ direction, while the
average precision of the track impact
parameter is $\sim 40 \mu$m.
A Gaussian fit to the
reconstructed invariant mass distribution for
signal events gives a resolution of 18 MeV/$c^2$.
The efficiency to identify
real muons from B decays is $\sim 95\%$, while the
probability to misidentify a hadron either
due to the occupancy in muon chambers or 
because it decays in flight is below $1\%$
for hadrons with momenta larger than
10 GeV/$c$. 
The combined trigger efficiency, 
for signal events passing the selection used in the analysis, is
greater than $90\%$.

%
The analysis is based
on a very efficient soft preselection
that removes a good fraction of the background while keeping most of
the signal falling within the LHCb acceptance \cite{mumunote}. 
Each event is then weighted by its
likelihood ratio on the relevant distributions. 
Three likelihoods were defined: 
a geometrical likelihood that takes into
account variables related to the vertex,
pointing and isolation; the muon 
identification likelihood; and the invariant
mass likelihood.

In Fig.~\ref{confidencelevel} (left), the 90$\%$ CL exclusion region for the
branching ratio is shown as a function of the integrated luminosity,  
under the assumption that only the expected background
is observed.
The two lines correspond to two background
hypothesis (i.e., nominal and shifted values), 
and the region between them represents the uncertainty coming from the
limited MC statistics.
LHCb has the potential to exclude the interesting region between $10^{-8}$
and the SM prediction with very little luminosity ($\sim 0.5
fb^{-1}$).

In Fig.~\ref{sensitivity} (right), the LHCb potential to discover a signal is
shown as a function of the integrated luminosity.
LHCb has the potential for a $3 \sigma$ ($5 \sigma$) observation
(discovery) of the SM prediction with $\sim 2\ \mathrm{fb}^{-1}$
($\sim 6\ \mathrm{fb}^{-1}$)
of data.

\section{Radiative decays $\mathrm{b \rightarrow s}\gamma$}
\label{sec:2} 
The radiative penguin b $\rightarrow$ s$\gamma$ decay is another
example of remarkable interest. Its total branching
fraction is very sensitive to physics beyond
the SM as it may be affected by the presence of
charged Higgs or SUSY particles in the loop. 
Presently, the world average 
is in good agreement with the theoretical SM prediction.
Nevertheless, in the SM the emitted photon in radiative 
decays is expected to be predominantly left-handed; this SM prediction 
is still untested, and 
right-handed components arise in a variety of new physics models.
No clarifying results have been obtained up to now due to limited statistics.

In principle, a test of the Standard Model can be made by
measuring the direct CP violation that results in a difference of the 
decay rates $\mathrm{B \rightarrow X} \gamma$ and 
$\mathrm{\bar{B} \rightarrow \bar{X} }\gamma$.
In SM the direct CP asimmetry is reliably predicted to be less than $1 \%$;
however in some SM extensions the contribution from new particles in the loop could
increase it up to $10 \% - 40\%$ \cite{radiativeasymmetries}.
Unfortunatly, inclusive decays are well described theoretically but are difficult to access experimentally;
while exclusive cases are theoretically much more difficult to calculate.

A more sensitive test of the SM 
can be made by
measuring the CP asymmetries from the interference of
mixing and decay amplitudes in radiative B neutral decays when B$^0_s$ and $\mathrm{\bar{B}^0_s}$
are required to have transitions to the same final state $X^0 \gamma$. If the photon
is polarized, as predicted in the SM, the CP asymmetry from the mixing should vanish \cite{mixingvanish}.
The current world average for this CP asymmetry is consistent with
0, but the errors are still large. 

In the LHCb experiment, radiative $\mathrm{b \rightarrow s}\gamma$ decays
can be reconstructed in the modes $\mathrm{B^0 \rightarrow K^{*0}} \gamma$ and 
$\mathrm{B^0_s} \rightarrow \phi\gamma$ \cite{radiativenote}. 
The main source of background is assumed to be $b \bar{b}-$inclusive events where at least
one b-hadron is emitted in the LHCb acceptance region.
The reconstruction algorithms
and the offline selection criteria
for the decays $\mathrm{B^0} \rightarrow \mathrm{K^{*0}} \gamma$, $\mathrm{K^{*0} \rightarrow K^+}\pi^-$
and $\mathrm{B^0_s} \rightarrow \phi\gamma$,  $\phi
\rightarrow \mathrm{K^+K^-}$ are similar.
Charged tracks have to be consistent with
the requested particle identification and inconsistent
with originating from the reconstructed
primary vertex. Selected K$^{*0}$ or $\phi$ candidates are
combined with photon candidates of transverse
energy larger than 2.8 GeV, in order to remove
low energy $\gamma$ and $\pi^0$. The reconstructed B
candidate is required to be compatible with coming
from a primary vertex, which is a very powerful cut against
combinatorial background. Finally, to suppress the correlated
background from the decays $\mathrm{B^0 \rightarrow K}^{*0}\pi^0$
and $\mathrm{B^0_s} \rightarrow \phi \pi^0$, a cut on the K$^{*0}$ and
$\phi$ decay (helicity) 
angle with respect to the B direction
is applied. The mass distributions of B candidates after the trigger
and off-line selection 
are expected to be $(69.9 \pm 2.2)\ \mathrm{MeV/}c^2$ and  
$(70.9 \pm 2.1)\ \mathrm{MeV/}c^2$ respectively.
The expected annual (2 fb$^{-1}$) yields and background over signal
ratios ($B/S$) 
are given in Tab.~\ref{yieldstable}. 
\begin{table}
\caption{Annual yields and background-to-signal ratios for selective decays at LHCb. $B/S$ ratios
are limited by MC statistics.}
\label{yieldstable}       
\begin{tabular}{lll}
\hline\noalign{\smallskip}
Mode & Yield & $B/S$   \\
\hline\noalign{\smallskip}
\hline\noalign{\smallskip}
B meson decays\\
\noalign{\smallskip}\hline\noalign{\smallskip}
$B^0 \rightarrow K^{0*} \gamma$ & 68000 & $<0.6$ \\
$B^0_s \rightarrow \phi\gamma$ & 11500 & $<0.55$ \\
\hline\noalign{\smallskip}
\hline\noalign{\smallskip}
B baryon decays\\
\noalign{\smallskip}\hline\noalign{\smallskip}
$\mathrm{\Lambda_b} \rightarrow \mathrm{\Lambda} \gamma$ & 750 & $<42$ \\
\hline\noalign{\smallskip}
$\mathrm{\Lambda_b} \rightarrow \mathrm{\Lambda}(1520) \gamma$ & 4200 & $<10$ \\
$\mathrm{\Lambda_b} \rightarrow \mathrm{\Lambda}(1670) \gamma$ & 2500 & $<18$ \\
$\mathrm{\Lambda_b} \rightarrow \mathrm{\Lambda}(1690) \gamma$ & 2200 & $<18$ \\
\hline\noalign{\smallskip}
\hline\noalign{\smallskip}
Electroweak decays\\
\noalign{\smallskip}\hline\noalign{\smallskip}
$B^0 \rightarrow \mu \mu K^0_s$ & 18774 & $\sim$ 3\\
$B^0 \rightarrow ee K^0_s$ & 9240 & $\sim$ 5\\
\noalign{\smallskip}\hline
\end{tabular}
\end{table}

\section{$\mathrm{\Lambda_b} \rightarrow \mathrm{\Lambda} \gamma$ polarization measurements}

Radiative decays of polarized $\mathrm{\Lambda_b}$ baryons 
to $\mathrm{\Lambda_b} \rightarrow \mathrm{\Lambda} \gamma$
represent an attractive possibility
to measure the helicity of the photon emitted
in the b $\rightarrow$ s quark transition \cite{lambdaref}.
The photon polarization can be tested by measuring the angular distribution 
of the photon in the $\mathrm{\Lambda_b}$ decay or even through the
angular distribution of the proton coming from the 
$\mathrm{\Lambda \rightarrow p}\pi^-$ decay. 

The study of this channel is challenging because the long lifetime of $\mathrm{\Lambda}$
baryon means that it will typically traverse a large fraction of the
tracking system before decaying. A possible solution is to 
consider decays to heavier $\mathrm{\Lambda}$ resonances; 
the subsequent decay to $\pi \mathrm{K^-}$ allows to 
trace back the decay of the $\mathrm{\Lambda_b}$.
The event selection is similar to the one presented above for B mesons \cite{lambdanote}. 
The expected event yields and $B/S$ ratios for 2 fb$^{-1}$ are given in Tab.~\ref{yieldstable}.
It can be noted that the heavier $\mathrm{\Lambda}$ modes are have a
higher statistical power, but 
since the distribution of the proton polarization is expected to be flat 
(i.e. the proton asymmetry is uniform $\alpha_p = 0$ due to parity
conservation), the intrinsic sensitivity is lower.
The main conclusion is that, assuming a $\mathrm{\Lambda_b}$
polarization of at least 20\%, LHCb 
can measure the right-handed component of the photon polarization  from
$\mathrm{\Lambda_b \rightarrow  \Lambda(1115)}\gamma$ 
decays down to 15\% at 3$\sigma$ significance after five years of
running. 
The additional
contribution from the $\mathrm{\Lambda(X)}$ resonances to the measurable range has been estimated
to be 2\% at most. 
The dependence of the photon polarization sensitivity on the initial
$\mathrm{\Lambda_b}$ polarization (in the range $P_{\mathrm{\Lambda_b}} = 20 - 100 \%$) has been found to be of the order of a few
percent.

\section{$A_{FB}$ measurement}
\label{sec:3} 

Electroweak $\mathrm{b \rightarrow s\ell\ell}$ decay is a FCNC process which proceeds via a 
$b \rightarrow s$ transition through a penguin diagram. New Physics processes can therefore
enter at the same level as SM processes.\\
In particular the branching ratio as a function of the squared invariant mass 
of the dilepton system can be 
affected in most New Physics scenarios.\\
However, the experimentally accessible exclusive decays are affected theoretically by
hadronic uncertainties. A possible solution is to study ratios where hadronic 
uncertainties are significantly reduced.\\
The forward--backward asymmetry
$A_{FB}$ is defined for the transition b $\rightarrow s\ell\ell$s 
by the angle $\theta_{\ell}$ between the $\ell^+$ and the
b hadron flight directions in the di-lepton rest
frame.
The shape of the asymmetry $A_{FB}$ as a
function of the lepton-lepton effective mass $m^2_{\ell\ell}$ and
especially the position of the zero crossing (i.e.
the $m^2_{\ell\ell}$ value corresponding to $A_{FB}$=0) are almost
unaffected by hadronic form factor uncertainties,
thus providing a good basis for searching
for deviations from the SM predictions \cite{semileptonicafb}.\\
Thanks to its very clean experimental signature, the exclusive decay 
$\mathrm{B_d \rightarrow K}^*\mu\mu$ has been chosen to extract $A_{FB}$.
The selection is based on the identification of two muons with opposite charge
and of the relevant hadronic final state \cite{afbnote}. Very strict requirements on
the vertex quality are applied to
reduce the backgrounds from cascade semileptonic b 
$\rightarrow \mu\nu$c, c $\rightarrow \mu\nu$s  and from two
semileptonic b $\rightarrow \mu\nu$c decays.
These processes have to be well under control, as they can induce a
bias on $A_{FB}$. 
The background from $\mathrm{c \bar{c}}$ 
resonances is removed by vetoing the J$/\psi$ and $\psi(2S)$
mass windows in the di-muon effective mass distribution. 

LHCb expects a 15 MeV/$c^2$ resolution on the B mass and 10 MeV/$c^2$
on the di-muon mass.
The resolution for $\theta_{FB}$ is 4 mrad.
The expected yield for one nominal year of running at LHCb (2 fb$^{-1}$) is about 7200 events
with a background to signal ratio $B/S \simeq 0.5$.
The overall trigger and reconstruction efficiency is estimated to be around $1 \%$.

\begin{figure}
\includegraphics[width=0.45\textwidth,angle=0]{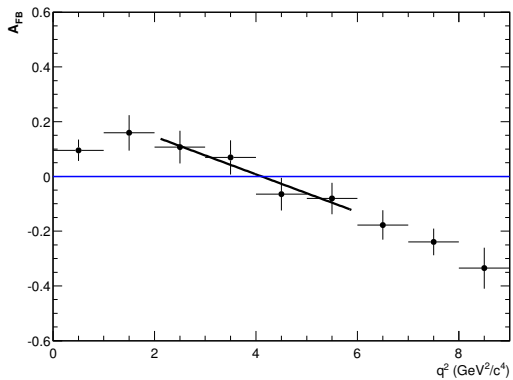}
\caption{The $A_{FB}$ distribution from an example 2 fb$^{-1}$ toy experiment. The solid
line shows the fit result.}
\label{AFB}       
\end{figure}

Using the results obtained from the full simulation of the $\mathrm{B^0 \rightarrow K}^{0*} \mu \mu$ channel,
LHCb has estimated the sensitivity to the forward-backward asymmetry in a "toy" MC study.
The typical behaviour of $A_{FB}$ versus $m^2_{\mu \mu}$, after one year of running at the 
nominal luminosity (2fb$^{-1}$) is shown in Fig.~\ref{AFB}.
With 2 fb$^{-1}$ of data the precision on the point of zero-crossing
$(A_{FB} = 0)$ is expected to be 0.46 GeV$^2/c^4$;
while at 10 fb$^{-1}$ the precision improves to 0.27 GeV$^2/c^4$.
No bias in the mean of the measured zero-crossing point is observed.

Recent theoretical work \cite{FL} has highlighted other interesting asimmetries to be 
studied, as the longitudinal polarization fraction of the $\mathrm{K}^{*0} (F_L)$ and the
second of the two polarization amplitude asymmetries ($\mathrm{A^{(2)}_T}$).
The parameters are all predicted with high precision from theory in the SM and many extensions
beyond the SM. The available statistics are limited by the requirement to restrict
the region of the di-muon masses from 1 to 6 GeV$^2$, which is favoured by
small theoretical errors.
In this region, the LHCb expected resolution with an integrated luminosity
of 2 fb$^{-1}$ is 0.016 in $\mathrm{F_L}$ and 0.42 in $\mathrm{A^{(2)}_T}$ \cite{FLnote}.

\section{$R_\mathrm{K}$ measurements at LHCb}
Finally,
the ratio of b $\rightarrow$ $\mu\mu$s and b $\rightarrow$ ees decays in any
exclusive mode is also a clean probe of the SM.
Lepton universality predicts this ratio to be unity
with a theoretical error below 1\% \cite{mumustoees}.
In the SM, the ratio of 
b $\rightarrow \mu\mu$s and b $\rightarrow$ ees decays is expected 
to be very close to unity, namely $R_\mathrm{K} = 1.000 \pm 0.001$.
Deviations of the order of 10\% can occur with 
neutral Higgs boson exchange in models that
distinguish between lepton flavours (for istance, the minimal SUSY
model at large tan$\beta$).

In LHCb the reconstruction of the two decay modes $\mathrm{B^+ \rightarrow K^+ \mu\mu}$ and 
$\mathrm{B^+ \rightarrow K^+ ee}$ allows an extraction of the ratio
$R_\mathrm{K}$ of the two branching fractions, 
integrated over a given di-lepton mass range \cite{rknote}.
The two decays are reconstructed with the same procedure and
requirements described
above, except that a proper brems-strahlung correction is essential in the $\mathrm{B^+ \rightarrow K^+ ee}$ channel.
The di-lepton mass range is chosen to be $1 < m^2_{ll} < 6$ GeV$^2/c^4$
in order to avoid $\mathrm{c \bar{c}}$
resonances. The event yields are extracted from a two-dimensional fit to the $\mathrm{K} \ell\ell$ and $\ell\ell$ masses
in order to take into account the backgrounds from b $\rightarrow J/\psi$ s and $\mathrm{B}^0 \rightarrow \mathrm{K}^{0*} \ell \ell$.
The expected yields are given in Tab.~\ref{yieldstable}.
With 10 fb$^{-1}$ we expect a relative error on $R_\mathrm{K}$ between 4\% and 6\% depending on the
level of background and the efficiency of the trigger. The study of
the most likely sources of
systematic errors shows that this error will be statistics-dominated.

\section{Conclusions}
The LHCb experiment has a promising potential for the study of rare
loop-induced decays, which are sensitive to new physics in many
Standard Model extensions. 
In particular, for the very rare decay $\mathrm{B^0_s} \rightarrow \mu
\mu$, 
present experiments will detect a signal only when the BR is strongly enhanced
by New Physics. With a sensitivity exceeding the BR expected in the SM, LHCb
will be able to discover both enhancements and suppression.
In addition, LHCb has good potential for measuring the helicity of the
photon emitted in the 
$\mathrm{b \rightarrow s} \gamma$ decay, the forward-backward asymmetry $A_{FB}$ for the transition 
b$~\rightarrow ll$s and the ratio of b~$\rightarrow \mu \mu$s and
b $\rightarrow$ ees decays in a number of
exclusive modes.

The experimental strategies, the expected annual signal event yields
and the estimates on background
to signal ratios have been presented.  

\label{sec:4}

%
\bibliography{}
%

%
%

\end{document}